# Accurate estimation of microscopic diffusion anisotropy and its time dependence in the mouse brain

Andrada Ianuş[a,b], Sune N. Jespersen[c,d], Teresa Serradas Duarte[a], Daniel C. Alexander[b], Ivana Drobnjak[b], Noam Shemesh[a*]

[a]*Champalimaud Neuroscience Programme, Champalimaud Centre for the Unknown, Lisbon, Portugal,*

[b]*Centre for Medical Image Computing, University College London, London, UK*

[c]*Center of Functionally Integrative Neuroscience (CFIN), Clinical Institute, Aarhus University, Aarhus, Denmark*

[d]*Department of Physics and Astronomy, Aarhus University, Aarhus, Denmark*

Abbreviations: µA – microscopic anisotropy, µFA – microscopic fractional anisotropy; SDE – single diffusion encoding; ODE – oscillating diffusion encoding; DDE – double diffusion encoding; DODE – double oscillating diffusion encoding; CNS – central nervous system; MRI – magnetic resonance imaging; dMRI – diffusion MRI; DTI – diffusion tensor imaging; MD – mean diffusivity; FA – fractional anisotropy; QSI – q-space imaging; DSI – diffusion spectrum imaging; DKI – diffusion kurtosis imaging; QTE – q-space trajectory encoding;

Running Head: Accurate mapping of microscopic anisotropy

*Corresponding author
Address: Champalimaud Neuroscience Programme, Champalimaud Centre for the Unknown, Av. Brasilia 1400-036, Lisbon, Portugal.
*Email:* noam.shemesh@neuro.fchampalimaud.org

**Abstract**

Microscopic diffusion anisotropy (µA) has been recently gaining increasing attention for its ability to decouple the average compartment anisotropy from orientation dispersion. Advanced diffusion MRI sequences, such as double diffusion encoding (DDE) and double oscillating diffusion encoding (DODE) have been used for mapping µA, usually using measurements from a single b shell. However, the accuracy of µA estimation vis-à-vis different b-values was not assessed. Moreover, the time-dependence of this metric, which could offer additional insights into tissue microstructure, has not been studied so far. Here, we investigate both these concepts using theory, simulation, and experiments performed at 16.4T in the mouse brain, ex-vivo. In the first part, simulations and experimental results show that the conventional estimation of microscopic anisotropy from the difference of D(O)DE sequences with parallel and orthogonal gradient directions yields values that highly depend on the choice of b-value. To mitigate this undesirable bias, we propose a multi-shell approach that harnesses a polynomial fit of the signal difference up to third order terms in b-value. In simulations, this approach yields more accurate µA metrics, which are similar to the ground-truth values. The second part of this work uses the proposed multi-shell method to estimate the time/frequency dependence of µA. The data shows either an increase or no change in µA with frequency depending on the region of interest, both in white and gray matter. When comparing the experimental results with simulations, it emerges that simple geometric models such as infinite cylinders with either negligible or finite radii cannot replicate the measured trend, and more complex models, which, for example, incorporate structure along the fibre direction are required. Thus, measuring the time dependence of microscopic anisotropy can provide valuable information for characterizing tissue microstructure.

# Introduction

Diffusion MRI (dMRI) probes the displacement of water molecules inside the tissue and can provide a unique window into cellular architecture at subvoxel dimensions. Thus, dMRI became highly applicable for studies of disease that alter tissue microstructure, such as stroke [1, 2], multiple sclerosis [3], Alzheimer's disease [4], etc., as well as for studies of brain plasticity [5] or development [6, 7], where changes in the microstructure precede gross anatomical variations. The most commonly used dMRI technique for brain studies is diffusion tensor imaging (DTI) [8], which assumes that the diffusion process is probed in the (anisotropic) Gaussian regime and reports metrics such as mean diffusivity (MD), fractional anisotropy (FA) and fibre direction. Although widely used in clinical applications, it is clear that the underlying microstructure is too complex to be fully characterized by a single diffusion tensor [9]. Various techniques aiming to overcome the limitations of DTI have been proposed in the literature. Approaches such as q-space imaging (QSI) [10, 11] or diffusion spectrum imaging (DSI) [12] have been developed to recover various higher-order properties of the diffusion process, while methods such as diffusion kurtosis imaging (DKI) [13] directly quantify the leading deviation from Gaussian diffusion. Other techniques aim to relate various tissue features, such as neurite density and orientation distribution [14-16][R1.8], axon diameter [17-21], membrane permeability [22, 23], to the diffusion signal and then solve the inverse problem to estimate parameters of interest.

Restricted diffusion induces a time-dependence of the diffusion tensor, which can be used as an additional source of information into the underlying tissue microstructure. Time- and frequency-dependencies of the diffusion coefficient have been studied in porous media [24] as well as in biological systems [25-28]. Several theoretical frameworks relate time-dependent behaviours to specific morphological features, such as pore size [29-32] as well as the internal disorder and packing [27, 33, 34]. Oscillating Diffusion Encoding (ODE) can be used to probe short time scales, and have demonstrated superior tensor contrasts [35], as well as sensitivity to surface-to-volume ratio [36, 37] and restriction size in elongated pores [38-40].

By contrast with these techniques, in which orientation- and size-distributions are difficult to disentangle, estimation of microscopic anisotropy (µA) provides a different measure of the restricting geometry, which can report on its anisotropy irrespective of the overall organization (e.g., orientation dispersion) on the voxel scale [41]. Thus, µA reflects more accurately microscopic tissue properties compared to the standard fractional anisotropy derived from DTI, and can be used as a potentially valuable biomarker. Single diffusion encoding (SDE) techniques can make various assumptions in order to model and quantify µA [42-44], but their constraints are not necessarily compatible with data acquired with different types of encoding, which can lead to biased quantification [45]. Moreover, in substrates with unknown microstructure, SDE acquisitions struggle to discriminate various microstructural configurations, such as randomly oriented anisotropic pores from distributed pore sizes [32, 41, 46-50].

To resolve this ambiguity, advanced diffusion acquisitions with varying gradient orientation in one measurement are advantageous [32, 41, 47, 49-51]. Double diffusion encoding (DDE) is now perhaps the most well-established approach for quantifying µA [41, 52, 53] from measurements performed using two independent pulsed gradient vectors that probe the correlation of water diffusion in different directions. DDEs are mostly used in the long mixing time regime, to ensure independence of the spin displacements within each compartment during the first and second encoding periods. In completely randomly oriented systems, theoretical studies predicted that such an approach can report on µA directly from the signal modulation [41, 54, 55], and this has been validated in systems such as phantoms [46], ex-vivo tissues and cells [56, 57], in-vivo rodents [57-59], and humans [60], as well as for clinical applications in multiple sclerosis [61]. Very recent advances in MR Spectroscopy have been able to detect DDE modulations for brain metabolites, thereby revealing their µA and confinement [59] and imparting sensitivity towards cell-specific neuronal and astrocytic microstructures [62]. To make the measurement rotationally invariant, several acquisition schemes have been proposed, mainly the 15-direction scheme by Lawrenz and Finsterbusch [63] and its subsequent extensions [64], and the DDE 5-design, which has been shown to provide even more accurate µA metrics [65]. To remove the dependence of µA on compartment size, normalized metrics of microscopic fractional anisotropy (µFA) were also reported [64, 65]. Q-space trajectory encoding (QTE) is another promising technique capable of delivering analogues of µFA in clinical imaging [50, 51, 66, 67], however, the standard QTE analysis assumes Gaussian time-independent diffusion, which might bias the estimated metrics [67].

With few exceptions [57], the DDE techniques described above are usually used to probe microscopic anisotropy at a fixed diffusion time. For instance, most DDE studies were performed with long diffusion and mixing times. Nevertheless, further insight into tissue architecture can be gained by varying diffusion times. Recent work has combined the DDE and ODE sequences in a double oscillating diffusion encoding (DODE) sequence [68], which employs two independent trains of oscillating gradients that can have different orientations. Thus, such an acquisition can be used to probe the frequency dependence of microscopic anisotropy. Additionally, one major advantage of DODE predicted by [68], is that the mixing time dependence effectively vanishes for most pore sizes, thereby facilitating the sequence's fulfilment of the long mixing time regime (required for µA analyses) for most practical acquisitions. Indeed, a recent study showed that µFA derived from DODE, but not DDE, correlates best with axon diameter and myelin content. Although quantification of µA for such sequences can be easily adapted from DDE, current approaches are based on sequences with a single b-value and assume that higher order terms $O(b^3)$ in the signal difference are negligible, which can affect the accuracy of the estimated metrics.

In the first part of the present study, we show both in simulation and measured data that quantification of µA is extremely sensitive to the choice of b-value, resulting in biased estimates. We then propose a multi-shell estimation scheme which accounts for higher order terms in the signal difference to provide accurate µA values. In the second part, we use the proposed multi-shell approach with DDE and DODE sequences to investigate the time/frequency dependence of microscopic anisotropy in the ex-vivo mouse brain. The patterns emerging from time-dependent µA are then shown to provide insights into the diffusion models which can describe tissue microstructure.

## Background and theory

*DDE and DODE sequences*

To investigate the dependence of microscopic anisotropy in the mouse brain on acquisition parameters, specifically diffusion weighting and time/frequency, we employ DDE and DODE sequences with different timing parameters, which are schematically illustrated in Figure 1a) and b), respectively. The DDE sequences are parametrized by pulse duration $\delta = \delta_1 = \delta_2$, diffusion time $\Delta = \Delta_1 = \Delta_2$, separation time $\tau_s$ (time interval between the two pairs of gradients; the corresponding mixing time is $\tau_s + \delta$), gradients amplitudes $G = G_1 = G_2$ and directions $\hat{g}_1$ and $\hat{g}_2$ chosen according to the 5-design scheme from [65] in order to provide a powder averaged signal. The 5-design with 12 parallel and 60 orthogonal measurements, ensures a rotationally-invariant quantification of µA.

The DODE sequences employ cosine-like trapezoidal waveforms described by pulse duration $\delta = \delta_1 = \delta_2$, number of half oscillation periods $N = N_1 = N_2$, separation time $\tau_s$ (time interval between the two gradient waveforms) as well as gradient amplitude and direction defined in the same way as for DDE. The oscillation frequency of the DODE waveforms is calculated as $\nu = N/2\delta$. The b-values for all the sequences are calculated according to the expressions derived in [69] which take into account the finite rise time of the gradient.

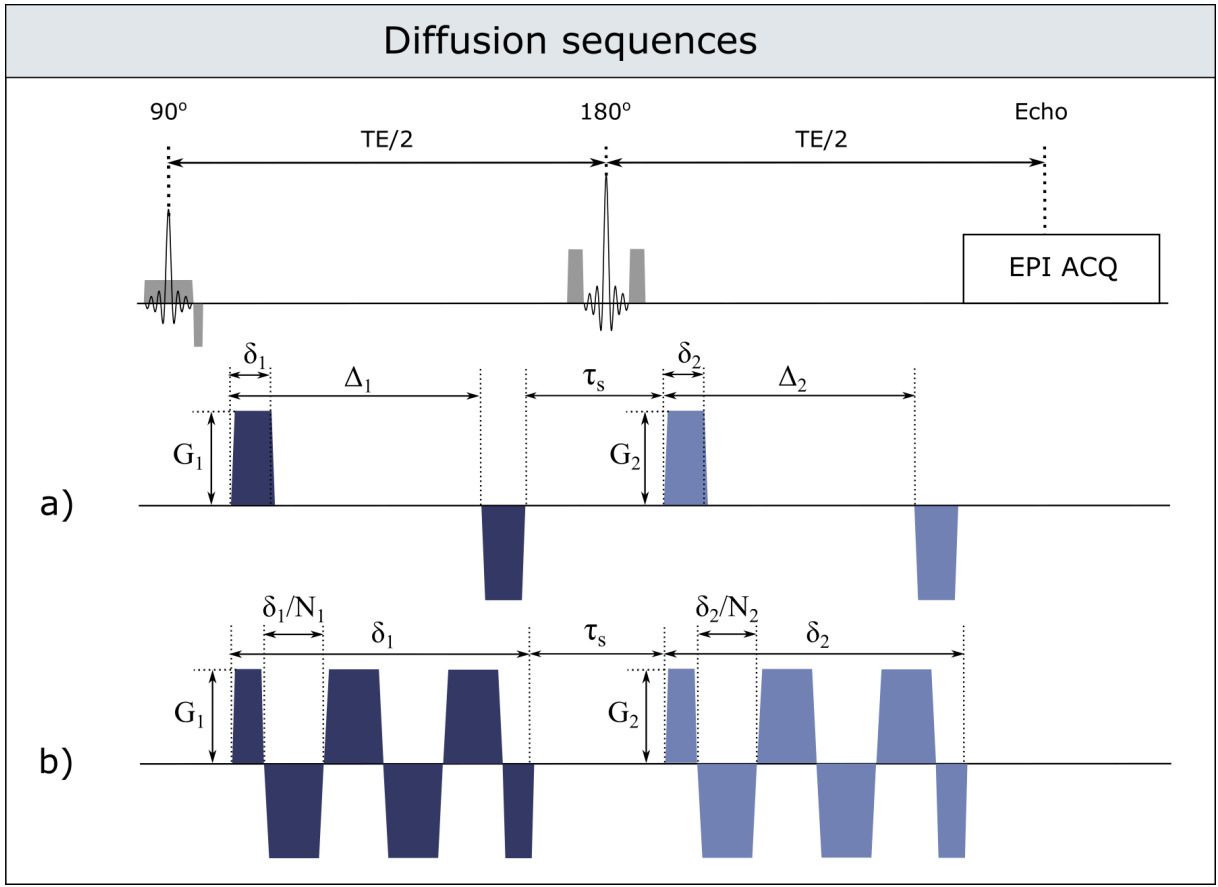

*Figure 1 Schematic representation of a) DDE and b) DODE diffusion sequences, with $N_{1,2}$ >1. DODE sequences with $N_{1,2}$ = 1 have two lobes and are equivalent with DDE sequences with gradient duration and diffusion time of $\delta_{DODE}/2$. For DDE sequences, $\delta$ is the gradient duration from ramp up until ramp down[R1.4], while for the DODE sequences, the total waveform duration is $\delta$ + rise time, to allow for the apodisation described in [70][R1.17]. The rise time of the gradient is 0.1 ms for all waveforms.*

## Quantification of microscopic diffusion anisotropy

The square of microscopic diffusion anisotropy, as defined in [53], is proportional to the variance over single pore diffusion tensor eigenvalues $\sigma_i$, i.e. $\mu A^2 \propto \text{var}(\sigma_i), i = \{1,2,3\}$. For very short diffusion times $\mu A^2$ is vanishing, while for diffusion times long enough to allow the spins to probe the entire pore space, $\mu A^2$ times diffusion time squared essentially becomes the pore eccentricity.

For DDE sequences with long mixing time, microscopic anisotropy can be derived from the difference of the averaged signals acquired with parallel ($S_\parallel$) and perpendicular ($S_\perp$) gradient directions [65]:

$$\mu A^2 = \frac{1}{b^2}\left[\log\left(\frac{1}{12}\sum S_\parallel\right) - \log\left(\frac{1}{60}\sum S_\perp\right)\right] \qquad (1)$$

where the average is computed over different gradient directions and $b = \gamma^2 G^2 \delta^2(\Delta - \delta/3)$, with γ the gyromagnetic ratio. For populations of identical pores, $\mu A^2$ is equivalent to $\frac{3}{5}\text{var}(\sigma_i)$.

For DODE sequences, we use a similar rationale and derive the expression of $\mu A^2$ based on the difference between DODE measurements with parallel and orthogonal gradients. To this end, we assume a diffusion model consisting of randomly oriented axially symmetric microdomains with frequency dependent parallel and perpendicular diffusivities ($D_\parallel(\omega)$ and $D_\perp(\omega)$), and we follow the derivation in [71, 72] to compute the powder averaged signal. For DODE sequences with parallel and perpendicular gradients, the expressions are the following:

$$S_\parallel^{p.a.} = \frac{1}{2}\int_0^\pi \exp\left(-(b_1 + b_2)(D(\omega)_\parallel \cos^2\theta + D(\omega)_\perp \sin^2\theta)\right)\sin\theta d\theta \qquad (2)$$

and

$$S_\perp^{p.a.} = \frac{1}{4\pi}\int_0^{2\pi}\int_0^\pi \exp[-b_1(D(\omega)_\parallel \cos^2\theta + D(\omega)_\perp \sin^2\theta) \\ - b_2(D(\omega)_\parallel \sin^2\theta \cos^2\varphi \\ + D(\omega)_\perp \sin^2\varphi + D(\omega)_\perp \cos^2\theta \cos^2\varphi)]\sin\theta d\theta d\varphi, \qquad (3)$$

where $b_1$ and $b_2$ are the b-values of the first and second gradient waveforms and the angles θ and φ define the gradient directions relative to each microdomain, and p.a. stands for powder average. Calculating the cumulant expansion up to second order in b, the difference between DODE sequences with parallel and perpendicular gradients when $b_1 = b_2$ is:

$$\log(S_\parallel^{p.a.}) - \log(S_\perp^{p.a.}) = b^2 \frac{2}{15}(D(\omega)_\parallel - D(\omega)_\perp)^2. \qquad (4)$$

Thus, $\mu A^2$ for a DODE sequence can be computed as:

$$\mu A^2 = \frac{\log(S_\parallel^{p.a.}) - \log(S_\perp^{p.a.})}{b^2} = \frac{2}{15}(D(\omega)_\parallel - D(\omega)_\perp)^2 = \frac{3}{5}\text{var}(\sigma_i), \qquad (5)$$

and is analogous to the expression for a DDE sequence described in equation (2).

*Higher order effects in quantification of µA*

In the previous analysis, the computation of $\mu A^2$ is based on the second order cumulant expansion of the signal and is usually evaluated at a single b-value, which might introduce a bias when the higher order terms are not vanishing. To correct for this contribution, we can expand equation (5) to include the next order terms ($b^3$):

$$\log\left(S_\parallel^{p.a.}(b)\right) - \log\left(S_\perp^{p.a.}(b)\right) = \mu A^2 b^2 + P_3 b^3, \quad (6)$$

where $\mu A^2$ denotes the corrected microscopic diffusion anisotropy metric computed from multi-shell data and $P_3$ reflects the contribution of 3rd order terms. For the substrate described above consisting of identical microdomains with time dependent diffusivities

$$P_3 = -\frac{8}{315}\left(D(\omega)_\parallel - D(\omega)_\perp\right)^3. \quad (7)$$

From here onwards we denote the apparent microscopic anisotropy measured from single shell data at a given b-value as $\widetilde{\mu A}^2$.

*Normalized microscopic anisotropy metric*

A convenient way to represent microscopic anisotropy and to remove its dependence on compartment size is to normalize it with respect to the size of the diffusion tensor. Thus, in analogy to macroscopic fractional anisotropy, the microscopic counterpart $\mu FA^2$ can be calculated as:

$$\mu FA^2 = \frac{3}{2}\frac{\mu A^2}{\mu A^2 + \frac{3}{5}MD^2}, \quad (8)$$

where $MD$ is the mean diffusivity of the diffusion tensor fitted to the D(O)DE data acquired with parallel gradient orientations. When data from multiple b-shells is used, diffusion and kurtosis tensors are fitted and MD is calculated from the eigenvalues of the diffusion tensor. The 3/2 factor ensures the same normalization as in the standard definition of FA [73].

## Methods

*Diffusion simulations*

The first part of this work investigates in simulation the dependence of estimated microscopic anisotropy on the b-value as well as on the timing parameters of the DDE and DODE sequences. We simulate the signal for protocols with identical timing parameters to the experimental ones presented in Table 1b) and b-values between 250 and 4000 s/mm$^2$, and various geometric models featuring microscopic anisotropy. For simulations, we use the MISST toolbox [74, 75], which implements a 3D extension of the multiple propagator framework. To reduce the model parameter space, we compute the powder averaged signal by simulating isotropically oriented microcompartments, thereby removing any directional information. For the geometric models, we use the nomenclature in [76].

We simulate signals for models consisting of anisotropic compartments widely used in the literature that feature Gaussian diffusion, such as AstroZeppelins (isotropically oriented cylindrically symmetric diffusion tensors) and AstroSticks (isotropically oriented sticks with unidimensional diffusion), as well as restricted diffusion, such as AstroCylinders (isotropically oriented infinite cylinders). Furthermore, to increase the complexity of the geometric models, we also consider sticks with finite lengths and a mixture of AstroSticks and Spheres. For Zeppelins, we simulate combinations of parallel and perpendicular diffusivities between 0.05 and 2 µm$^2$/ms, while for cylinders we simulate combinations of radii between 0.25 and 5 µm and lengths between 5 to 50 µm, with a diffusivity value of 2 µm$^2$/ms.

In the first analysis, for the microstructural models described above, we investigate the dependence of apparent $\widetilde{\mu A}^2$ on b-value for DODE sequences with δ = 10 ms, N = 4 and b values between 250 and 4000 s/mm$^2$. Then, we correct for the effect of higher order terms by fitting equation (6) to the signal difference between measurements with parallel and perpendicular gradients. Further, we compare the corrected microscopic anisotropy $\mu A^2$ with the ground-truth value, which is calculated in the following way: first, diffusion and kurtosis tensors are fitted to the signal from each pore separately using the measurements with parallel gradients and all b-values, then $\mu A^2$ is calculated for each pore from the variance of the DT eigenvalues ($\frac{3}{5}\text{var}(\sigma_i)$) and finally, it is averaged over different pore sizes and orientations in each substrate.

In the second simulation, we analyse the dependence of mean diffusivity as well as the corrected microscopic anisotropy metric $\mu A^2$ for different diffusion sequences and substrates, which we compare with the ground-truth values calculated as described above for each sequence.

a)

| | | | Water phantom, 22° | | | |
|---|---|---|---|---|---|---|
| | Imaging parameters: | TE / TR (ms) | Matrix size | FOV (mm x mm) | In-plane resolution (mm x mm) | Slice thickness (mm) |
| | | 51.3 / 2700 | 84 x 76 | 10.1 x 9.1 | 0.12 x 0.12 | 2 |
| | DDE sequences: | b value (s/mm²) | δ (ms) | Δ (ms) | $\tau_s$ (ms) | Gradient directions |
| | | 1000 | 0.9 | 10, 5 | 16.5 | 5-design, positive and negative |
| | DODE sequences: | b value (s/mm²) | δ (ms) | N / ν (Hz) | $\tau_s$ (ms) | Gradient directions |
| | | 1000 | 15 | 2, 4, 6, 8 10 / 66, 133, 200, 266, 333 | 5 | 5-design, positive and negative |

b)

| | | | Mouse brain, ex-vivo, 37° | | | |
|---|---|---|---|---|---|---|
| | Imaging parameters: | TE / TR (ms) | Matrix size | FOV (mm x mm) | In-plane resolution (mm x mm) | Slice thickness (mm) |
| | | 52 / 3000 | 88 x 76 | 10.6 x 9.1 | 0.12 x 0.12 | 0.7 |
| | | | Experiment 1: b-value dependence | | | |
| | DODE sequences: | b value (s/mm²) | δ (ms) | N / ν (Hz) | $\tau_s$ (ms) | Gradient directions |
| | | 500, 750, 1000, … 3750, 4000 | 15 | 5 / 166 | 5 | 5-design, positive and negative |
| | | | Experiment 2: sequence dependence | | | |
| | DDE sequences: | b value (s/mm²) | δ (ms) | Δ (ms) | $\tau_s$ (ms) | Gradient directions |
| | | 1000, 1750, 2500, 3250, 4000 | 1.7 | 10, 5 | 16.5 | 5-design, positive and negative |
| | DODE sequences: | b value (s/mm²) | δ (ms) | N / ν (Hz) | $\tau_s$ (ms) | Gradient directions |
| | | 1000, 1750, 2500, 3250, 4000 | 15 | 2, 3, 4, 5, 6 / 66, 100, 133, 166, 200 | 5 | 5-design, positive and negative |

*Table 1 Imaging and diffusion parameters for a) water phantom and b) ex-vivo mouse brain acquisitions.*

*Experiments*

All experiments have been performed on a Bruker Aeon Ascend 16.4 T scanner interfaced with an Avance IIIHD console and equipped with gradients capable of producing up to 3000 mT/m in all directions, and controlled by Paravision 6.01. All DDE/DODE sequences were written in-house. All animal experiments in this study were preapproved by the local ORBEA committee for animal welfare and ethics, in accordance with Portuguese and EU laws.

Phantom validation

To validate the sequences' implementation, we performed tests in two phantoms. First, we used a phantom consisting of a 5 mm NMR tube filled with a mixture of $H_2O$ and $D_2O$ (1:4) doped with copper sulphate ($CuSO_4$). The acquisition details for the imaging parameters as well as DDE and DODE diffusion sequences used for the water phantom experiment are detailed in Table 1a). The images were acquired using single-shot EPI readout with a bandwidth of ~555 kHz and 1.20 partial Fourier. For each DDE and DODE protocol 8 non-diffusion weighted images (b = 0 s/mm$^2$) and two sets of the 72-direction diffusion weighted images were acquired, using the 5-design gradient orientations: one with the original orientation scheme, and another with inverted directions, so that cross-terms can be cancelled out [77]. The gradient strengths were adjusted to yield the specified b-value, having amplitudes between 0.32 and 1.59 T/m.

Second, to test that the protocols used for imaging the mouse brain do not yield any significant artifacts in this set-up, such as due to concomitant gradients (e.g. [78]), we used a phantom consisting of a 10 mm NMR tube filled with a solution of PVP40 (Polyvinylpyrrolidone, Sigma Aldrich, Lisbon, Portugal) with a mass concentration of 40% in a mixture of $H_2O$ and $D_2O$ (1:9), which has similar diffusivity to ex-vivo tissue. For this phantom we used a DODE imaging protocol with the same parameters as in Experiment 2 detailed in Table 1b, except for a slightly larger field of view (12.8 x 11 mm) to cover the sample, repetition time of 4 s, and 3 frequencies (N = 2, 4 and 6 / ν = 66, 133 and 200 Hz). The phantom was scanned at room temperature of 22 °C.[R1.1]

Ex-vivo mouse brain imaging

Specimen preparation: two brain samples[R1.3] were extracted from healthy adult mice weighing ~25 gr by standard intracardial PFA perfusion and preserved in a 4% PFA solution at 4°C. Before scanning, the brains were socked in phosphate buffered saline (PBS) for 24h and placed in a 10 mm NMR tube filled with fluorinert. All samples were scanned at 37°C.

The acquisition details for the imaging parameters as well as DDE and DODE diffusion sequences used for brain imaging are detailed in Table 1b). Experiment 1 was performed in one brain, while experiment 2 was performed in both brains[R1.3]. The images were acquired using single-shot EPI readout with a bandwidth of ~555 kHz and 1.20 partial Fourier. For each DDE and DODE protocol 8 non-diffusion weighted images (b=0 s/mm$^2$) and two sets of the 72-direction diffusion weighted images were acquired, using the 5-design gradient orientations with both positive and negative directions. The gradient strengths were adjusted to yield the specified b-value, having amplitudes less than 1580 mT/m, except for DODE with N=6 where the maximum gradient strength is 1890 mT/m. The SNR of the data is around 35 for b = 1000 s/mm$^2$, 22 for b = 2500 s/mm$^2$ and 15 for b = 4000 s/mm$^2$, and the acquisition took approximately 60h for one sample[R1.7]. To show the robustness of the measured trends, the second brain was placed upside down in the NMR tube[R1.3].

Data analysis

Pre-processing: brain images have been first denoised using the random matrix theory approach [70, 79] (with a kernel of size 11), then Gibbs ringing effects were removed using the unringing algorithm in [80]. Then, the geometric average for pairs of measurements with opposite gradient directions was computed to remove any effects of cross-terms with imaging gradients [77]. The second brain was registered slice by slice to the first one using the affine registration algorithm in Matlab®. Pre-processed data has been analysed using home-written code in Matlab® (The MathWorks Inc., Natick, MA, USA).

*Experiment 1 – b-value dependence and accurate extraction of microscopic anisotropy*

The aim of the first experiment is to investigate the dependence of apparent microscopic anisotropy $\widetilde{\mu A}^2$ on b-value in the mouse brain and to propose a multi-shell approach for accurate estimation of microscopic anisotropy. To correct for higher order signal contributions, we perform a polynomial fit to the signal difference in equation (6), fitting the coefficients of $b^2$ and $b^3$ terms. For this analysis, we use data acquired in one brain sample with DODE sequences with δ = 15 ms and N = 7 (ν = 166Hz), and 15 b-values linearly spaced between 500 and 4000 s/mm$^2$. For such an analysis, DODE sequences are preferable over DDE sequences, as the influence of the separation time on the signal is considerably smaller. This ensures that linear terms in b which would appear in the expression of the signal difference in equation (7) for short mixing times are indeed negligible [81].

*Experiment 2 - Time/frequency dependence of diffusion metrics*

The aim of the second experiment is to investigate the dependence of different diffusion metrics on the timing of the acquisition sequence, and was performed in two brain samples. Specifically, we focus on mean diffusivity (MD) and fractional anisotropy (FA), the corrected microscopic anisotropy ($\mu A^2$) and its normalized counterpart $\mu FA$. MD and FA were calculated from the eigenvalues of the diffusion tensor obtained when fitting the diffusion kurtosis model (DKI) [13] to data acquired with parallel gradient orientations and all the b-values for a given frequency. The DKI fit was performed using a non-linear least squares algorithm in Matlab®[R1.6]. For each sequence, $\mu A^2$ was calculated by fitting the polynomial expression in equation (6) to the signal difference measured at 5 different b-values, as described in Table 1b). The normalized $\mu FA$ was then computed using the corrected $\mu A^2$ values and the MD values. To investigate the dependence of these metrics on frequency, we perform a statistical analysis for voxels in four white matter ROIs (medial and lateral of corpus callosum, cerebral peduncle and internal capsule) and four gray matter ROIs (cortex, thalamus, piriform cortex, striatum) which have been manually delineated and are the same for both brains. For the statistical analysis, we use a random intercept and random coefficient mixed effects model where the relevant diffusion metric (measured at each voxel) is considered as the dependent variable and the frequency v as the explanatory variable. Thus, the variable v is nested in a voxel identifier, which is nested in a subject identifier. The significance level is adjusted for the number of voxels using the conservative Bonferroni correction[R1.2a].

## Results

*Simulations*

One of the questions this study aims to answer, is how accurate is the estimation of microscopic anisotropy, when measured at different b-values. Figure 2 plots the apparent microscopic anisotropy $\widetilde{\mu A}^2$ estimated from DODE experiments performed using a range of b-values and different models of microstructure featuring either Gaussian diffusion (Figure 2a, 2b and 2d) or restricted diffusion (Figure 2c, 2e and 2f) (blue curves), as well as the ground-truth values $\mu A^2_{g.t.}$ (yellow lines). For all models, clearly, the $\widetilde{\mu A}^2$ estimated from a single b-value strongly depends on the specific b-value employed. We postulated that this dependence arises from contributions of higher-order terms in the signals. Indeed, when $\mu A^2$ is computed using the information from all b-values to correct for higher order terms, similar values to the ground-truth are obtained. Slight departures are present for substrates with a mixture of sizes. Moreover, for most substrates with similarly sized pores, the estimated P$_3$ coefficient is in good agreement with its ground-truth value computed in a similar way to $\mu A^2_{g.t.}$, (less than 14% difference for zeppelins, sticks, cylinders and finite sticks), while for zeppelins with a distribution of diffusivities and the mixture of sticks and spheres, where the assumption of identical pores fails, the difference is larger, i.e. 20% and 39%, respectively.

The second objective of this study was to investigate the time/frequency dependence of μA. Prior to probing this question with experiments, we sought to gain insight from further simulations. Figure 3 plots the corrected microscopic anisotropy metrics, $\mu A^2$ and the corresponding ground-truth values $\mu A^2_{g.t.}$, as well as the estimated MD, as a function of the timing parameters of DDE (to probe different times) and DODE (to probe different frequencies) for microstructural models featuring restricted diffusion. For the AstroSticks model, and other models featuring Gaussian diffusion (not shown), microscopic anisotropy $\mu A^2$ and MD do not depend on diffusion time/frequency. For the AstroCylinders model (Figure 3a and 3b) $\mu A^2$ decreases, while MD increases with frequency. When investigating pores of finite length, as well as a mixture of sticks and spheres, the time/frequency dependence becomes more complex. For AstroFiniteSticks (Figure 3c and d), both $\mu A^2$ and MD overall increase with frequency, while for AstroFiniteCylinders (data not shown) $\mu A^2$ increases less. For the sticks and spheres model considered here, $\mu A^2$ increases with decreasing diffusion time and then plateaus for higher frequencies, and MD also increases with frequency. Due to the finite pulse length, the power spectra of DODE sequences, i.e. the squared Fourier transform of the diffusion gradient time integral, are not ideal with a sharp peak at the given frequency, but also have secondary peaks and harmonics which influence the observed frequency dependence, as illustrated in Figure 3h. There is a good agreement between the estimated $\mu A^2$ and the ground-truth values, especially for DODE sequences with higher frequencies, as the separation time becomes much larger than the characteristic diffusion time and terms linear in b are negligible, as assumed in the derivation of $\mu A^2$.

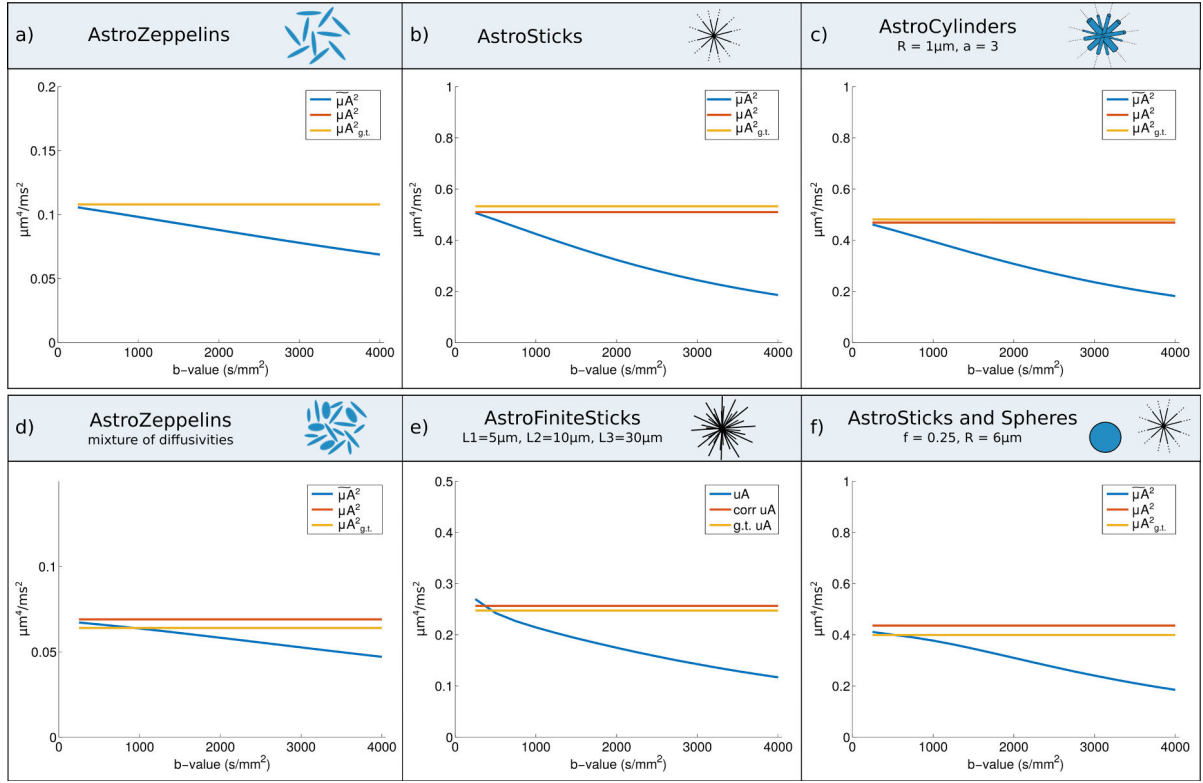

*Figure 2 Apparent microscopic anisotropy ($\widetilde{\mu A^2}$) as a function of b-value for different microstructural models as well as the corrected anisotropy metric ($\mu A^2$) and the ground-truth values ($\mu A^2_{g.t.}$). The parameters used for the substrates are the following: a) AstroZeppelins (isotropically oriented cylindrically symmetric tensors with $D_{||}$ = 1 µm²/ms and $D_\perp$ = 0.1 µm²/ms); b) AstroSticks (isotropically oriented sticks with $D_{||}$ = 2 µm²/ms); c) AstroCylinders (isotropically oriented cylinders with D = 2 µm²/ms and Gamma distributed radii with a mean of 1 µm[R1.5a] and a shape parameter of 3); d) AstroZeppelins with a mixture of diffusivities ($D_{||}$ = {0.5, 1, 1} µm²/ms, $D_\perp$ = {0.1, 0.1, 0.5} µm²/ms and corresponding volume fractions of 0.2, 0.5 and 0.3, respectively); e) AstroFiniteSticks (isotropically oriented sticks with an equal mixture of lengths L = {5, 10, 50} µm); f) AstroSticks and Spheres (isotropically oriented sticks and spheres with radius of 6 µm and a volume fraction of 0.25).*

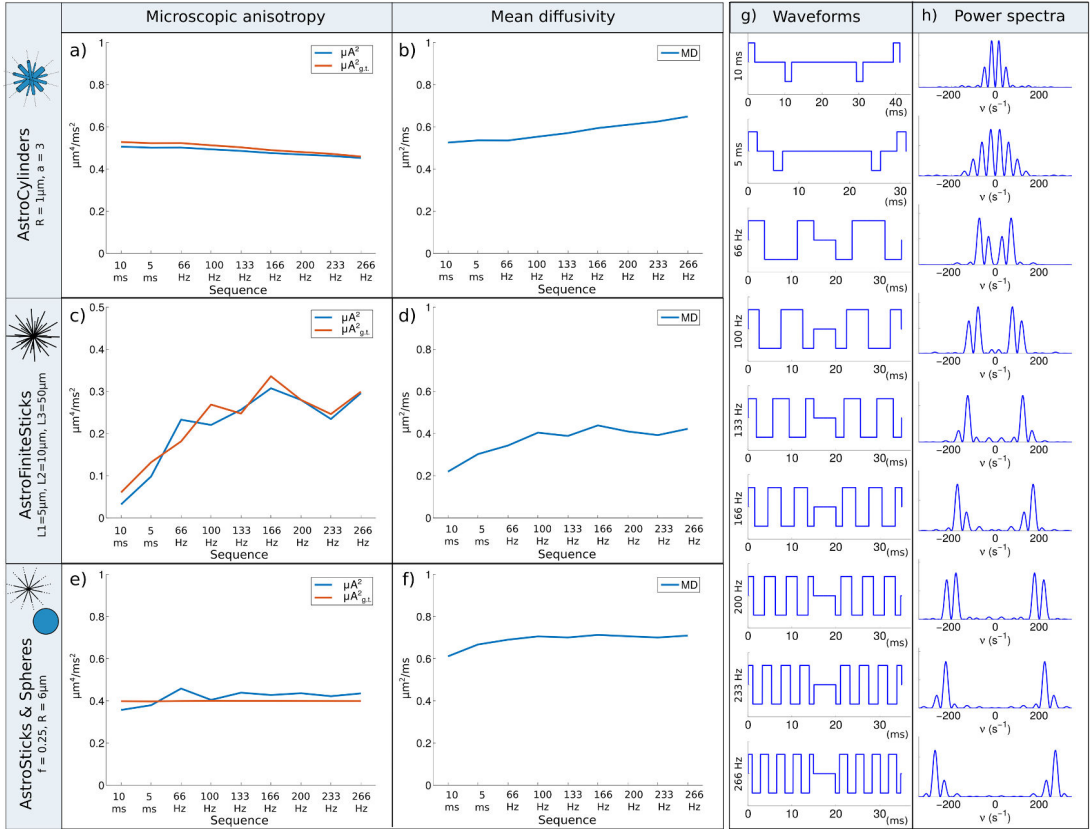

*Figure 3 a), c), e) Corrected anisotropy metric (μA$^2$) and the ground-truth values (μA$^2_{g.t.}$) as a function of the sequence timing parameters. b), d), f) Mean diffusivity as a function of sequence timing parameters. The first two points on the x-axis represent DDE sequences with two different Δ/τ$_s$ combinations, and the rest of the points represent DODE sequences with different frequencies. The microstructural models have the same parameters as the equivalent ones in Figure 2 c), e) and f) Schematic representation of the diffusion gradient waveforms and their corresponding power spectra.*

## Phantom validation

We then sought to study experimentally the predictions of the simulations above and the new DODE sequences presented here for the first time were validated on two phantoms.

For the water phantom, Figure 4a and 4b show the raw data for non-weighted and diffusion weighted (DODE, N = 10) images, while Figure 4c plots the mean diffusivity (MD) map for the same sequence. Figure 4d presents the estimated MD for different acquisitions, validating that it does not depend on the timing parameters of the sequences, as expected for free diffusion. The average MD value is 2.11±0.02 μm$^2$/ms, in agreement with the diffusivity of water at 22°C and off-the-shelf DTI experiments which yielded an MD value of 2.15±0.04 μm$^2$/ms. Figure 4e illustrates the difference between measurements with parallel and perpendicular gradient directions, which is negligible for all sequences. These results show that DDE and DODE sequences have been properly implemented, have the correct b-values, and have no artifacts for the gradient strengths used here (< ~1.6 T/m).

The experiments in the PVP phantom aimed to show that there were no artifacts in the estimation of the diffusion metrics using almost the same acquisition as in the mouse brain. Figure 4f shows the parameter maps for MD, FA, $\mu A^2$ and µFA$^2$ derived from the full protocol with N = 6, following the analysis described for the theory section. Figure 4g shows the median and interquartile range of these metrics for 3 frequencies (66, 133 and 200 Hz). There is no statistically significant change with frequency in any of the metrics, following the analysis outlined in the methods section. The MD values measured with DODE sequences (0.50±0.03 µm$^2$/ms) overlap with those estimated from an "off-the-shelf" DTI acquisition (0.51±0.08 µm$^2$/ms) and are similar to values previously reported in the literature [82]. $^{R1.1}$

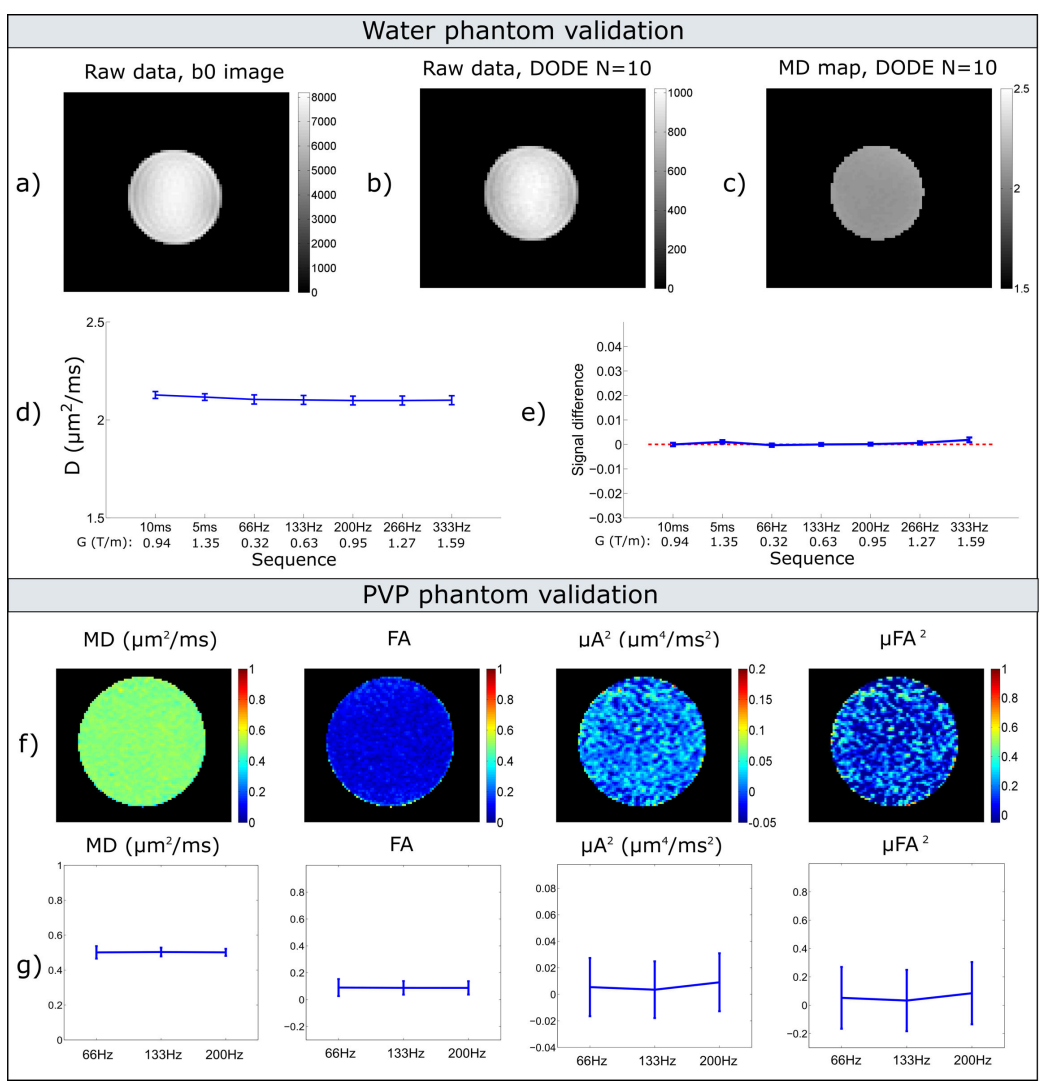

*Figure 4 Water phantom results: raw data for a) non-diffusion and b) diffusion weighted (DODE, N = 10 / 333 Hz, orthogonal gradients) images; c) MD map calculated for DODE sequences with N = 10; d) Estimated MD and e) signal difference between measurements with parallel and perpendicular gradients for different DDE and DODE sequences. The first two points on the x-axis represent DDE sequences with two different Δs, and the rest of the points represent DODE sequences with different frequencies; the gradient strength is also reported for each sequence. The dashed line in Figure 4e) represents the zero mark. PVP phantom results: f) maps of diffusion metrics (MD, FA, µA², µFA) calculated from the DODE protocol with N = 6; g) median and interquartile range of diffusion metrics for three different frequencies (v = 66, 133 & 200 Hz)[R1.1].*

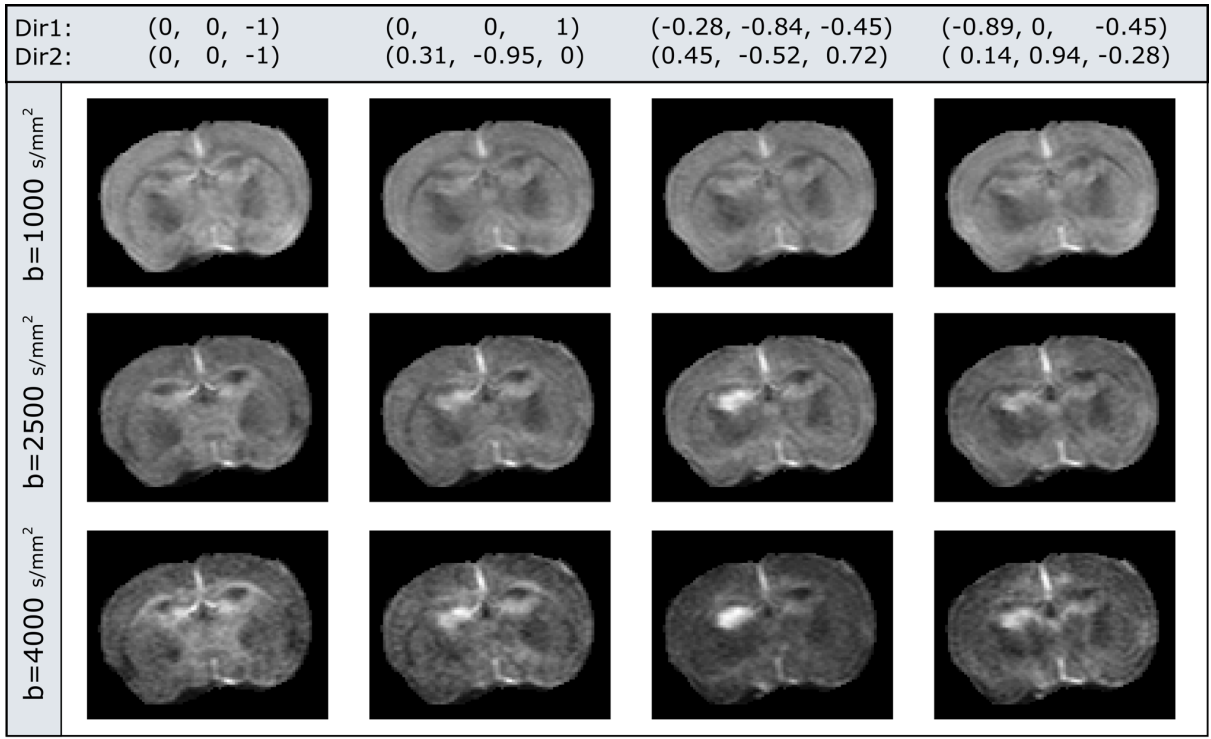

*Figure 5 Example raw data for DODE sequences with N=5 at three different b-values and four different gradient orientations.*

## Experiment 1 - b-value dependence of microscopic anisotropy

To test whether a similar dependence of microscopic anisotropy on b-values would emerge also in neural tissue, we employ the DODE dataset with a frequency of 200 Hz acquired for 15 b-values between 500 and 4000 s/mm². Figure 5 presents example raw data with three different b-values and four different gradient directions. Figure 6 illustrates $\widetilde{\mu A}^2$ maps measured at each b-value using the DODE dataset. The plots show indeed that $\widetilde{\mu A}^2$ values decrease with increasing b-value, with a more pronounced dependence in white matter. Moreover, the maps derived from data acquired at low b-values (< 1000 s/mm²) are very noisy, as the difference between measurements with parallel and perpendicular gradients is very small, and thus the effect of noise gets amplified.

Figure 7 presents the corrected microscopic anisotropy map $\mu A^2$, as well as the fitted polynomial coefficient ($P_3$) corresponding to the third order term in b in equation (7). $\mu A^2$ values are higher compared to the $\widetilde{\mu A}^2$ values measured at larger b values (> 2000 s/mm²), which are usually employed in DDE studies. The $P_3$ map shows that the strongest decrease with b-value is present in white matter, while in gray matter the $P_3$ values are closer to zero.

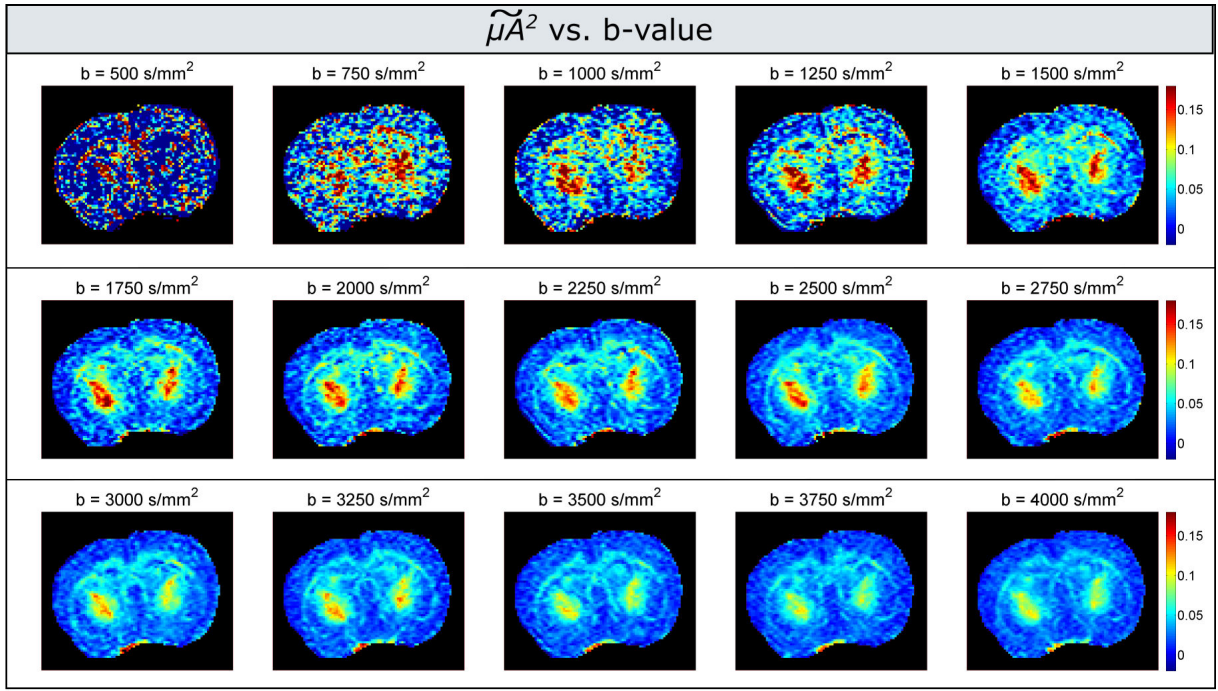

*Figure 6 Apparent microscopic anisotropy maps ($\widetilde{\mu A}^2$) for DODE sequences with N = 5 (166 Hz) and b values between 500 and 4000 s/mm².*

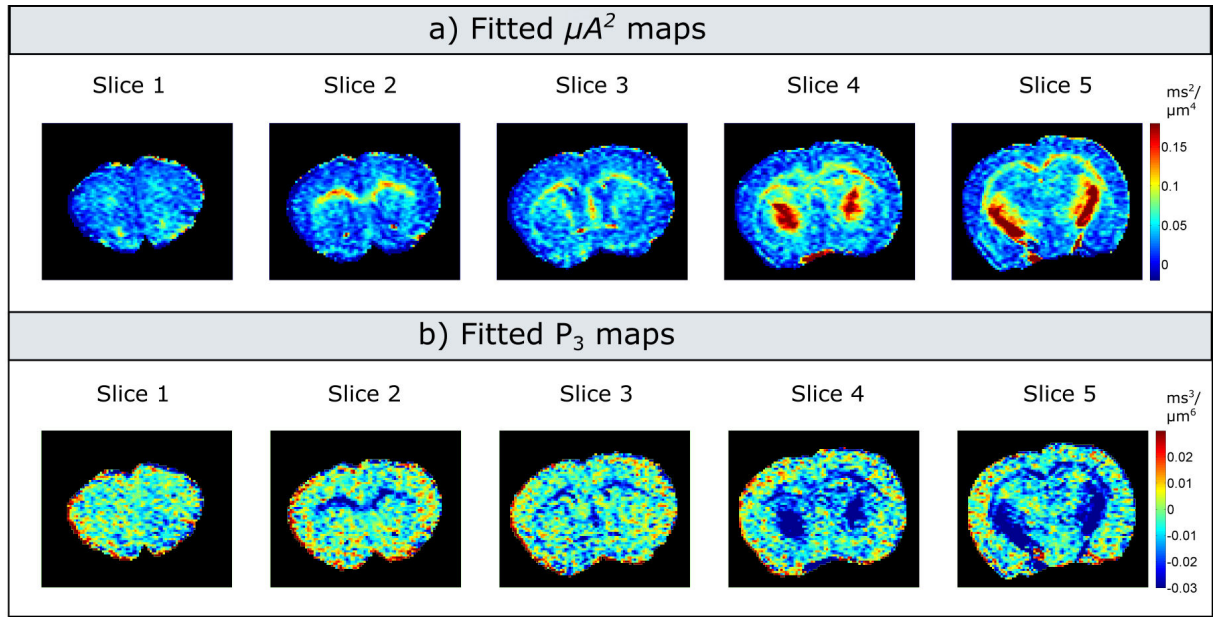

*Figure 7 a) Microscopic anisotropy maps calculated using the mutli-shell approach ($\mu A^2$) and b) corresponding polynomial coefficient map ($P_3$) for the $b^3$ terms in equation (7) calculated for DODE sequences with N = 5 (166 Hz).*

*Experiment 2 - Time/frequency dependence of diffusion metrics*

After ensuring that microscopic anisotropy can be assessed accurately using our novel multi-shell approach, we sought to investigate time/frequency dependencies of microscopic anisotropy. Figure 8 illustrates representative parameter maps for MD, FA, $\mu A^2$ and $\mu FA$ for DDE and DODE with different timing parameters in the first brain sample, and similar patterns were obtained in the second sample as well[R1.3]. The results show that MD increases with frequency, while FA slightly decreases with frequency. $\mu A^2$ slightly increases with frequency, while $\mu FA$ increases in some regions and decreases in others.

A more quantitative description of time/frequency dependence can be assessed using ROI analysis for the frequency dependence of various metrics. Figure 9a and 9b illustrate the choice of ROIs in gray matter (cortex, thalamus, piriform cortex and striatum) and white matter (medial and lateral corpus callosum, cerebral peduncle and internal capsule) and the dependence of MD, FA, $\mu A^2$ and $\mu FA$ on the timing parameters of DDE and DODE sequences. The median and interquartile range of the diffusion metrics shown in Figure 9 are computed over ROI voxels pooled from both brain samples[R1.3]. Table 2a summarises the results of the statistical analysis which tests the dependence of the diffusion metrics on the frequency of DODE sequences. The slope characterizing the change of the diffusion metrics with frequency is given for different ROIs, and the darker shaded cells represent statistically significant values[R1.2a]. The results confirm that MD increases with frequency (slopes between $0.43 \times 10^{-3}$ and $0.54 \times 10^{-3}$ µm²/ms/Hz in gray matter and between $0.48 \times 10^{-3}$ and $0.92 \times 10^{-3}$ µm²/ms/Hz in white matter), while a small FA decrease with frequency is significant in most ROIs considered here except for piriform cortex, and cerebral peduncle. $\mu A^2$ values are significantly higher for the DDE sequence with shorter diffusion compared to the other DDE sequence in most gray matter ROIs (slopes between -0.001 and -0.0015 µm⁴/ms²/s), while in white matter there is a significant difference only for the media corpus callosum (-0.0027 µm⁴/ms²/s). For DODE sequences, there is a significant increase in $\mu A^2$ with frequency in most gray and white matter ROIs, except for cerebral peduncle (slopes between $0.06 \times 10^{-3}$ and $0.1 \times 10^{-3}$ µm⁴/ms²/Hz in gray matter and between $0.15 \times 10^{-3}$ and $0.26 \times 10^{-3}$ µm⁴/ms²/Hz in white matter). When considering the normalized microscopic anisotropy metric $\mu FA$, the dependence on frequency is more variable, with a significant increase in some gray matter

ROIs (cortex and piriform cortex), a significant decrease in cerebral peduncle and no significant change in the other ROIs. $\mu FA$ values, which directly reflect the microscopic anisotropy of tissue without the effect of fibre orientation, are significantly higher ($p<<0.01$) than FA values in all the ROIs considered here, for both DDE and DODE measurements. The relative difference between $\mu FA$ and $FA$ is larger in the gray matter compared to white matter, as illustrated in Table 2b. For white matter ROIs, the relative difference between $\mu FA$ and $FA$ is higher in the internal capsule and corpus callosum compared with the cerebral penduncle, which is consistent with the amount of fibre dispersion measured in previous studies [83]. Negative values in $\mu A^2$ can occur both due to noise, as well as due to the sequence not satisfying the long mixing time assumption, and corresponding $\mu FA$ values were set to 0 and not included in the frequency analysis. Nevertheless, similar trends and significance levels were observed for $\mu FA^2$ which included all voxels[R1.2b]. In general, the $\mu FA$ estimates are noisier in white matter compared to gray matter, as structures are smaller with fewer voxels in the given WM ROIs, and more prone to partial volume effects.

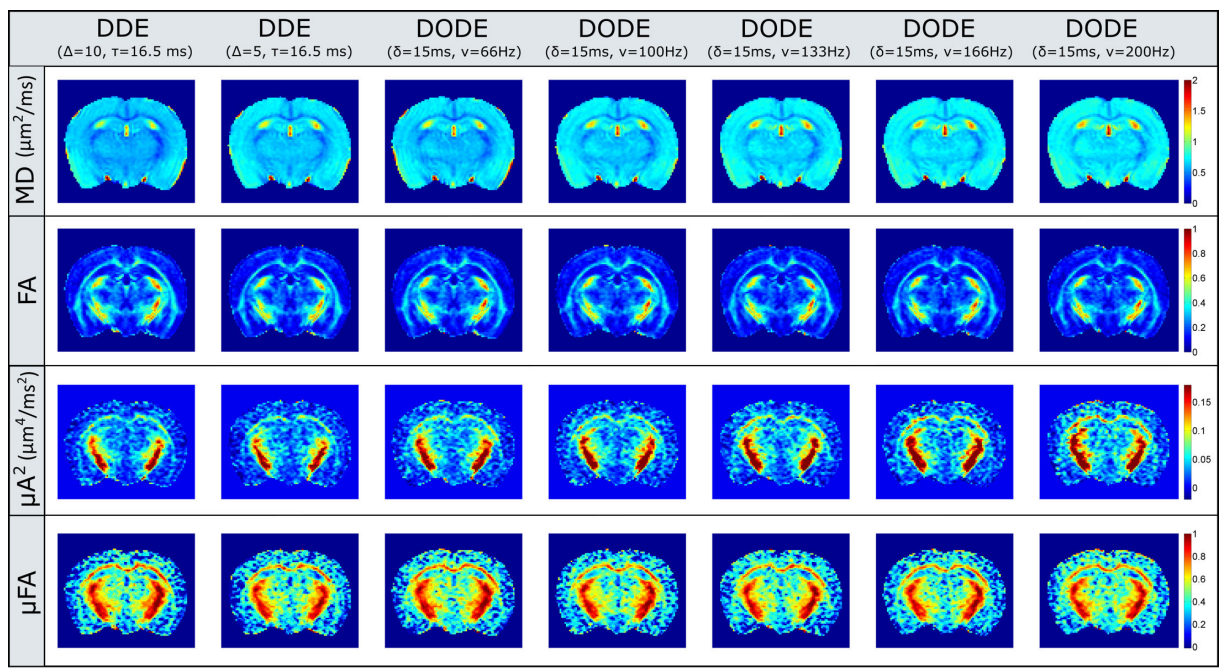

*Figure 8 Diffusion derived metrics (MD, FA, $\mu A^2$, μFA) for DDE and DODE sequences with different timing parameters.*

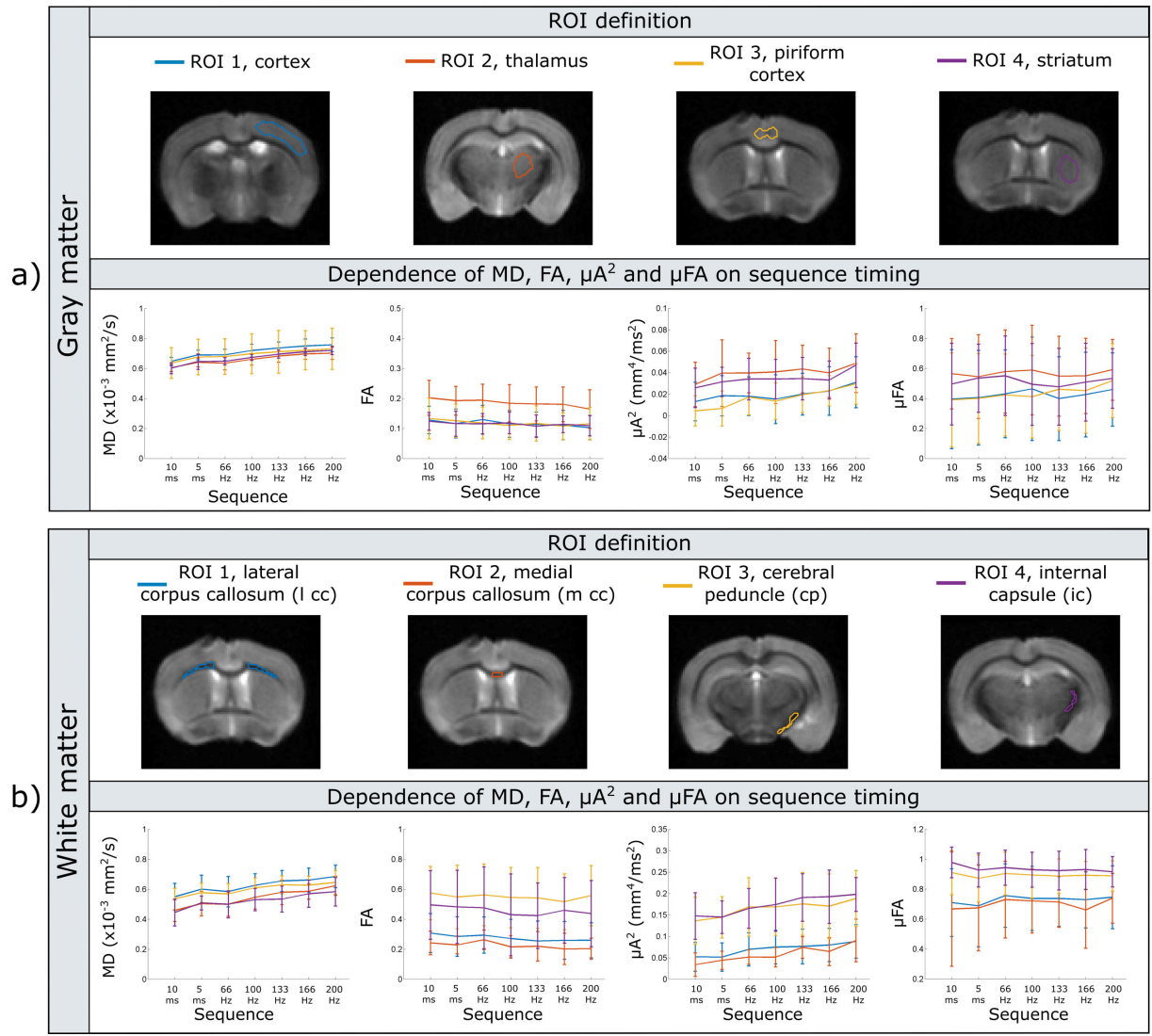

*Figure 9 Dependence of diffusion metrics (MD, FA, µA², µFA) on the timing parameters of DDE and DODE sequences for four ROIs in a) gray matter (cortex, thalamus, piriform cortex and striatum) and b) white matter (medial and lateral corpus callosum, cerebral peduncle and olfactory tracts). The plots show the median and interquartile range of the parameters computed over the ROI voxels from the two brain samples[R1.3] and each ROI is represented by a different colour.*

|   a) | Gray matter ROIs | | | | White matter ROIs | | | |
|---|---|---|---|---|---|---|---|---|
|   | ROI 1 | ROI 2 | ROI 3 | ROI 4 | ROI 1 | ROI 2 | ROI 3 | ROI 4 |
| $d\,MD/dv\,(\times 10^{-3})$ | 0.48 | 0.54 | 0.43 | 0.54 | 0.78 | 0.92 | 0.48 | 0.52 |
| $d\,FA/dv\,(\times 10^{-3})$ | -0.16 | -0.17 | -0.06 | -0.07 | -0.29 | -0.30 | -0.15 | -0.33 |
| $d\,\mu A^2/dv\,(\times 10^{-3})$ | 0.10 | 0.06 | 0.09 | 0.07 | 0.15 | 0.26 | -0.07 | 0.17 |
| $d\,\mu FA/dv\,(\times 10^{-3})$ | 0.27 | -0.13 | 0.42 | 0.05 | -0.16 | 0.03 | -0.23 | -0.21 |

| b) | | Gray matter ROIs | | | | White matter ROIs | | | |
|---|---|---|---|---|---|---|---|---|---|
| | | ROI 1 | ROI 2 | ROI 3 | ROI 4 | ROI 1 | ROI 2 | ROI 3 | ROI 4 |
| DDE 10 ms | FA | 0.13 | 0.20 | 0.13 | 0.12 | 0.30 | 0.30 | 0.57 | 0.50 |
| | µFA | 0.35 | 0.56 | 0.23 | 0.49 | 0.71 | 0.67 | 0.91 | 0.98 |
| DDE 5 ms | FA | 0.12 | 0.19 | 0.13 | 0.12 | 0.28 | 0.28 | 0.55 | 0.48 |
| | µFA | 0.38 | 0.54 | 0.31 | 0.54 | 0.69 | 0.67 | 0.87 | 0.92 |
| DODE 66 Hz | FA | 0.13 | 0.19 | 0.12 | 0.12 | 0.29 | 0.29 | 0.56 | 0.48 |
| | µFA | 0.38 | 0.58 | 0.40 | 0.54 | 0.75 | 0.73 | 0.90 | 0.94 |
| DODE 133 Hz | FA | 0.11 | 0.18 | 0.12 | 0.11 | 0.25 | 0.25 | 0.54 | 0.42 |
| | µFA | 0.38 | 0.54 | 0.43 | 0.47 | 0.74 | 0.71 | 0.88 | 0.92 |
| DODE 200 Hz | FA | 0.10 | 0.16 | 0.12 | 0.11 | 0.26 | 0.26 | 0.55 | 0.44 |
| | µFA | 0.45 | 0.59 | 0.50 | 0.53 | 0.75 | 0.74 | 0.89 | 0.91 |

Table 2 a) Slope estimated from the statistical model illustrating the dependence of various metrics (MD, FA, dµA², µFA) on the frequency of DODE sequences in different gray and white matter ROIs. Darker shaded cells represent statistically significant values ($p < 0.05$/number of voxels, adjusted for multiple comparisons using the conservative Bonferoni correction). b) Median FA and µFA values for different ROIs and acquisition sequences. The shaded cells colour code the relative difference (µFA-FA)/µFA in five intervals between 0.3 (lightest shade) and 0.8 (darkest shade)[R1.2a].

## Discussion

Mapping microscopic anisotropy using advanced diffusion MRI sequences, such as double diffusion encoding, provides a marker of tissue microstructure while mitigating the effects of orientation dispersion, and has been gaining popularity in neuroimaging studies. This work harnesses DODE and DDE acquisitions to study μA in the mouse brain, and its aims were two-fold: (1) to provide a multi-shell approach for accurate quantification of μA, and (2) to study its time/frequency dependence. In the first part, we show that standard single b-value quantification of μA results in biased estimates, and propose a method for obtaining an accurate estimation of μA which requires data samples from multiple b-values and a higher order fit. In the second part, we map the corrected μA metrics and perform a comprehensive characterization of their frequency dependencies in the mouse brain, using the advanced DODE sequence which was previously introduced theoretically in [68]. The main advantage of DODE is that it easily fulfils the long mixing time regime, which is highly advantageous for such characterizations. To our knowledge this and XXX are the first to use this pulse sequence, which we are happy to provide on request. Below, we elaborate and discuss each of these findings.

*Dependence of $\widetilde{\mu A}^2$ on the b-value*.

Nearly all previous studies on DDE have focused mainly on estimation of μA using a single b-value. Our simulations were designed to investigate μA in simple systems where the ground-truth is a-priori known, and the results clearly show that the estimated $\mu A^2$ decreases with b-value for a variety of microstructural models which feature microdomains with either Gaussian or restricted diffusion. Thus, measuring apparent $\widetilde{\mu A}^2$ at a single b-value, can bias the estimates compared to the ground-truth, especially for higher b-values (> 3000 s/mm$^2$). On the other hand, estimating $\widetilde{\mu A}^2$ from data acquired at low b-values (<1500 s/mm$^2$) results in very noisy estimations, as the difference between measurements with parallel and perpendicular gradients becomes comparable to the standard deviation of noise characteristic for most practical (and indeed, even state-of-the-art) DWI acquisitions. These trends were clearly shown in the experimental data, that was acquired with very good SNR at 16.4 T (~35 for b = 1000 s/mm$^2$, ~22 for b = 2500 s/mm$^2$ and ~15 for b = 4000 s/mm$^2$ after denoising) to avoid bias due to measurement noise. Although, the maps at low b-values show very noisy μA contrast, the b-value dependence was also clearly evident from the experimental results. This dependence corresponds to the simulation predictions, and requires the higher-order term correction to improve the accuracy of μA estimation.

*Accurate estimation $\mu A^2$ from multi-shell acquisitions*.

Once the bias in apparent $\widetilde{\mu A}^2$ became clear both from simulation and experiments and its origins traced to the higher order terms, we devised a correction scheme that would enable an accurate estimation of this important quantity. The simulations indicated that a model of the D(O)DE signal difference which includes both second and third order terms in b can be fitted to data acquired at multiple b-values to obtain a much more reliable estimate of microscopic anisotropy, which was found to be similar to the expected ground-truth value (c.f. Figure 2). For substrates which consist of identical microdomains, the corrected $\mu A^2$ estimates are almost identical to the ground-truth values, while small departures can be seen in substrates which feature a distribution of pore sizes/diffusivities. Again, this suggests that many μA metrics reported so far using data from a single b-value may have been underestimated, e.g. [64, 65]. When microscopic anisotropy is estimated from a data set acquired at a single b-value, then a compromise between SNR and estimation bias needs to be considered. In our experiments, a good balance was observed for data acquired at b-values between 2000 and 3000 s/mm$^2$.

It is important to note that Equation (7) assumes that terms linear in b are negligible. This implies that the long mixing time regime has been reached [84]. When this assumption is violated, the choice of mixing time can further bias microscopic anisotropy estimates from DDE sequences or DODE with the lowest frequencies, (Figure 3c) especially when size distributions are involved. In these cases, some of the pores may require longer times to reach the long $t_m$ regime, and these pores will also contain a large fraction of spins contributing to the signal. In such cases, linear terms in b are also present, and the interpretation of the b$^2$ coefficient as microscopic anisotropy is no longer as accurate. On the other hand, DODE sequences have been shown to be quite independent on the separation time between the two waveforms (for most practical experimental conditions), especially when $\tau_s$ is larger than the oscillation period of the gradient and its particular value does not have a significant effect on the power spectrum of the waveform [68]. In this case the assumption of negligible linear terms in b holds, and the estimates of microscopic anisotropy are closer to the ground-truth values (Figure 3a and f).

An alternative way to estimate microscopic anisotropy is to use the expressions of $\mu A^2$ and P3 derived in equation 5 and 7, respectively, and to fit only one variable, namely $D(\omega)_\parallel - D(\omega)_\perp$. For the simulations presented in Figure 3, computing $\mu A^2$ from the fitted $D(\omega)_\parallel - D(\omega)_\perp$ yields less accurate values, especially for substrates with restricted diffusion. In experimental data, applying this analysis to the DODE dataset with 15 b values yields maps that are less noisy and have smaller values compared to those in Figure 7 a. For the dataset containing DDE and DODE sequences with different timing parameters and 5 b values, the fitting fails in a certain number of voxels, mostly in the gray matter, where $\mu A^2$ is smaller and the effect of noise is more pronounced, leading to negative values that are not allowed. Overall, we found that fitting $\mu A^2$ and P3 separately provides more accurate and robust estimates, as it does not make any assumptions except for being in the long mixing time regime.

*Time/frequency dependence of $\mu A^2$.*

Time/frequency-dependencies in SDE have been proposed as fingerprints for different microstructural properties [25, 27]. However, the time/frequency dependence of microscopic anisotropy measured with D(O)DE sequences for different microstructural models has not been studied yet. Therefore, we first performed simulations, where the ground-truth is known a-priori. Indeed, the results show a different behaviour depending on the type of microstructural model analysed, as illustrated in Figure 3. For the simple stick model, we do not expect a time/frequency dependence, while for infinite cylinders, with either a single radius or a mixture of radii, microscopic anisotropy is expected to decrease with frequency. A similar trend is also observed using Monte Carlo simulations [85] in substrates featuring parallel cylinders with gamma distributed radii (mean radius 1μm, shape parameter 3) that include the effect of extracellular space. On the other hand, for models which include restriction along the fibre orientation, the time dependence of microscopic anisotropy is more complex and can show an increase with frequency. On the other hand, mean diffusivity increases with frequency in all the substrates featuring restricted diffusion. Thus, time-dependent measurements can potentially inform a choice of microstructure models which would best explain experimental data. The signal derivation for DODE sequences in equations 3)-6) has been explicitly written for ideal sequences with a delta power spectrum at frequency v. For the realistic power spectra depicted in Figure 3g) and h), the signal is calculated as the integral the integral over diffusion spectrum, i.e. $S = S_0 \exp(-\frac{1}{\pi} \int (F(\omega)D(\omega)F^*(\omega)\,d\omega)$ , where *F(ω)* is the Fourier transform of the time integral of the gradient waveform [30]. Nevertheless, the interpretation of µA computed from the powder averaged signal holds [65].

We then sought to test the actual time/frequency dependencies in the fixed mouse brain. Using five b-values, the corrected $\mu A^2$ was estimated, but more "conventional" metrics such as MD or FA were also extracted from the data. Consistent with previous studies using oscillating gradients [86], our results show that MD increases with frequency, while FA slightly decreases [86]. By contrast with MD and FA, the microscopic anisotropy metric $\mu A^2$, as well as its normalized counterpart $\mu FA$, showed more variable trends. $\mu A^2$ evidenced an increase with frequency in most ROIs, except for cerebral peduncle, while $\mu FA$ exhibited both increases (in cortex and piriform cortex) as well as decreases (in cerebral peduncle)[R1.2,1.3] with increasing frequency. A variation of $\mu FA$ between different white matter ROIs was observed, consistent with previous pre-clinical studies in in-vivo mouse brain [87], ex-vivo monkey brain [65], as well as diffusion tensor microimaging of the ex-vivo mouse brain [83]. Future work will also aim to establish which histological features are the most likely cause of the trends observed in this study.

The experimental data revealed an increase in MD and either an increase or no change in $\mu A^2$ with frequency. Comparing these results with the simulated trends suggests that tissue microstructure can be better explained by including structure along the fibre directions and/or restriction in close to isotropic pores, while simple infinite cylinders/sticks models cannot replicate the frequency dependence trends of MD and $\mu A^2$ observed in the brain. However, the jagged frequency dependence observed in simulations (Figure 3c), resulting from diffusion in pores with finite sizes, is not as preeminent in the experimental data, suggesting that diffusion along fibres is not necessarily fully restricted. This is also consistent with the conclusions of a recent DODE study in rat spinal cord. Moreover, contributions from extracellular space and the effect of noise can mask such small variations with frequency. A similar deduction can be made from a theoretical standpoint. Considering the simple case of a diffusion tensor, if $D(\omega)_\parallel$ is constant and only $D(\omega)_\perp$ increases with frequency, then both $\mu A^2$ and $\mu FA$ are decreasing. On the other hand, the trends we see in the data (i.e. an increase in both $\mu A^2$ and $\mu FA$ or a decrease in $\mu FA$ without a significant change in $\mu A^2$, combined with an increase in MD) can only result if an increase in $D(\omega)_\parallel$ with frequency also exists, suggesting some degree of structure along the parallel direction. A pronounced decrease in $D(\omega)_\perp$ is unlikely over the time range considered here, as compartment exchange times are probably much longer [88]. Overall, the observed trends are in agreement with recent observations [89], but perhaps in contrast with other recent studies probing cell-specific metabolites, which suggested an infinite stick model would be more appropriate for describing the microstructure [59, 90]. However, while those studies focused on cell-specific metabolites, this study lacks the specificity to a particular compartment due to water's ubiquity in all tissue environments, including extracellular spaces. Moreover, water and metabolites may also interact differently with the microstructure (e.g., in terms of permeability, diffusion constants, etc.), and thus water may effectively probe different environments compared with the cell-specific metabolites. Measurements from advanced sequences can also be used to investigate the validity and delineate different assumptions used in microstructure models, however, care needs to be taken when comparing the diffusion environment (ex-vivo/in-vivo) and the range of sequence parameters.

The current simulations investigated only a subset of simple geometries, and there are many other factors which could explain the trends observed in the experimental data. For instance, more complex geometries which include the effect of undulation [91, 92], neurite branching and/or the presence of spines [93]. Modelling the effects of membrane permeability could also affect the signal time dependence, however, for the diffusion times used in this study we expect a negligible effect, as the exchange times reported in the literature are an order of magnitude longer [22, 88]. Future work will investigate such effects.

This study covered a range of frequencies between 66 and 200 Hz and b-values up to 4000 s/mm$^2$, which were achieved using very strong gradients up to 1.9 T/m. Due to the fast T2 decay at ultra-high field (16.4 T) of ~20 ms [94], the gradient duration was limited to 15 ms, which in turn restricted the range of available frequencies. The time/frequency dependence can also be probed on more standard preclinical systems as well as the state-of-the-art Connectome human scanner. However, with limited gradient strength only lower frequencies can be probed while achieving the b-values desired for estimating $\mu A^2$ (2000 - 3000 s/mm$^2$).

Overall, the time/frequency dependence of microscopic anisotropy and other associated metrics measured with D(O)DE sequences probes an additional dimension of the diffusion process and can provide important information regarding the microscopic tissue architectures.

**Acknowledgements**

This study was supported by funding from the European Research Council (ERC) under the European Union's Horizon 2020 research and innovation programme (Starting Grant, agreement No. 679058), EPSRC (grant number M507970, M020533 and N018702), Danish National Research Foundation (CFIN), and The Danish Ministry of Science, Innovation, and Education (MINDLab). The authors are thankful to Dr. Daniel Nunes for preparing the brain specimens ands to Dr. Francesco Grussu for a useful discussion on statistical analysis.